%% file: main.tex
\documentclass[11pt]{article}
\usepackage[margin=1in]{geometry}
\usepackage{times}
\usepackage[
backend=biber,
style=alphabetic,
sorting=nyt,
maxnames=7,
maxalphanames=7,
backref=true,
url=false
]{biblatex}
\usepackage{xcolor}
\usepackage[colorlinks=true,citecolor=blue,linkcolor=blue]{hyperref}

\hypersetup{
    colorlinks,
    linkcolor={red!50!black},
    citecolor={blue!50!black},
    urlcolor={blue!80!black}
}

\usepackage[]{amsmath,amssymb,amsfonts,latexsym,amsthm,enumerate,fullpage,xcolor,bbm}
\usepackage{algorithm2e}
\usepackage[nameinlink, noabbrev, capitalize]{cleveref}
\usepackage{comment}
\usepackage{nicefrac}
\crefname{prop}{Proposition}{Propositions}
\crefname{ineq}{inequality}{inequalities}
\creflabelformat{ineq}{#2(#1)#3}

\newtheorem{theorem}{Theorem}
\newtheorem{lemma}{Lemma}

\newtheorem{claim}{Claim}

\newtheorem{definition}{Definition}

\crefname{THM}{Theorem}{Theorems}

\newtheorem{seccounter}{Counter}[section]
\newtheorem{sectheorem}[seccounter]{Theorem}
\newtheorem{seclemma}[seccounter]{Lemma}
\newtheorem{secproposition}[seccounter]{Proposition}
\newtheorem{secclaim}[seccounter]{Claim}
\newtheorem{secfact}[seccounter]{Fact}

\newtheorem{seccorollary}[seccounter]{Corollary}
\newtheorem{secdefinition}[seccounter]{Definition}

\usepackage{subcaption}
\usepackage{graphicx}
\usepackage{tikz}
\usepackage{color, colortbl}
\definecolor{LightCyan}{rgb}{0.88,1,1}
\definecolor{Gray}{gray}{0.9}
\usetikzlibrary{positioning}
\usetikzlibrary{decorations.pathmorphing}
\usetikzlibrary{automata,positioning}
\usetikzlibrary{arrows}
\usetikzlibrary{arrows.meta}
\usetikzlibrary{calc}

\usepackage{enumitem}
\newlist{casesenum}{enumerate}{2}
\setlist[casesenum,1]{label=\textbf{Case~\arabic*.}, 
  itemindent=*,leftmargin=0pt}
\setlist[casesenum,2]{label=\textbf{Case~\roman*.}, 
  itemindent=*,leftmargin=\parindent}

\newcommand{\E}{\mathbb{E}}
\newcommand{\N}{\mathbb{N}}

\newcommand{\F}{\mathbb{F}}

\newcommand{\poly}{\operatorname{poly}}

\newcommand{\supp}{\text{support}}
\newcommand{\pr}{{\prime}}

\newcommand{\U}{\mathbf{U}}
\newcommand{\X}{\mathbf{X}}
\newcommand{\Y}{\mathbf{Y}}

\newcommand{\A}{\mathbf{A}}
\newcommand{\B}{\mathbf{B}}

\newcommand{\RR}{\mathbf{R}}

\newcommand{\SSS}{\mathbf{S}}

\renewcommand{\multicitedelim}{\addsemicolon\space}

\newcommand{\zo}{\{0,1\}}

\newcommand{\Ext}{\mathsf{Ext}}

\newcommand{\sExt}{\mathsf{sExt}}

\newcommand{\Disp}{\mathsf{Disp}}

\newcommand{\eps}{\varepsilon}
\newcommand{\AC}{\mathsf{AC}}

\newcommand{\cF}{\mathcal{F}}
\newcommand{\cC}{\mathcal{C}}
\newcommand{\cX}{\mathcal{X}}

\newcommand{\cP}{\mathcal{P}}

\newcommand{\cH}{\mathcal{H}}
\newcommand{\minH}{H_\infty}

\usepackage{subfiles}
\newcommand{\dobib}{
    \printbibliography
}

\begin{document}
\renewcommand{\dobib}{}
\renewcommand*{\multicitedelim}{\addcomma\space}

\title{Extractors for Polynomial Sources over $\F_2$}
 \author{Eshan Chattopadhyay\thanks{Supported by a Sloan Research Fellowship and NSF CAREER Award 2045576.}\\ Cornell University\\ \texttt{eshan@cs.cornell.edu} \and Jesse Goodman\footnotemark[1]\\ Cornell University\\ \texttt{jpmgoodman@cs.cornell.edu} \and Mohit Gurumukhani\footnotemark[1] \\ Cornell University\\ \texttt{mgurumuk@cs.cornell.edu} }
\date{}

 \maketitle
 \begin{abstract}
 We explicitly construct the first nontrivial extractors for degree $d \ge 2$ polynomial sources over $\mathbb{F}_2^n$. Our extractor requires min-entropy $k\geq n - \tilde{\Omega}(\sqrt{\log n})$. Previously, no constructions were known, even for min-entropy $k\geq n-1$. A key ingredient in our construction is an \emph{input reduction lemma}, which allows us to assume that any polynomial source with min-entropy $k$ can be generated by $O(k)$ uniformly random bits.

We also provide strong formal evidence that polynomial sources are unusually challenging to extract from, by showing that even our most powerful general purpose extractors cannot handle polynomial sources with min-entropy below $k\geq n-o(n)$. In more detail, we show that \emph{sumset extractors} cannot even \emph{disperse} from degree $2$ polynomial sources with min-entropy $k\geq n-O(n/\log\log n)$. In fact, this impossibility result even holds for a more specialized family of sources that we introduce, called \emph{polynomial non-oblivious bit-fixing (NOBF) sources}. Polynomial NOBF sources are a natural new family of algebraic sources that lie at the intersection of polynomial and variety sources, and thus our impossibility result applies to both of these classical settings. This is especially surprising, since we \emph{do} have variety extractors that slightly beat this barrier - implying that sumset extractors are not a panacea in the world of seedless extraction.
 \end{abstract}

\maketitle


\subfile{sections/introduction}

\subfile{sections/prelims}

\subfile{sections/constructions}

\subfile{sections/impossibility}

\subfile{sections/open}

\subfile{sections/acknowledgements}

\printbibliography


\end{document}

%% file: sections/introduction.tex
\section{Introduction}



 Randomness is a very important resource in computation. It is widely used in theoretical and practical implementations of algorithms, distributed computing protocols, cryptographic protocols, machine learning algorithms, and much more \cite{motwani1995randomized}. Unfortunately, the randomness produced in practice is not of the highest quality and the corresponding distribution over bits is often biased and has various correlations \cite{herrero2017quantum}. To overcome this, an extractor is used to convert this biased distribution to a uniform distribution. The extractors used in practice are based on unproven theoretical assumptions and so the theoretical study of constructing efficient extractors is important. Extractors usually come in two flavors: seeded and seedless extractors. We focus here on the latter, and whenever we mention \emph{extractor} this is what we mean. 

Towards this end, let us formally define extractors for a class of distributions.

 \begin{definition}[Extractor]
A function $\Ext:\zo^n\rightarrow \zo^m$ is called an $\varepsilon$-extractor for a class $\cX$ of distributions over $\zo^n$ if for all \(\X\in\cX\),
\[
|\Ext(\X) - \U_m| \le \varepsilon,
\]
where \(|\cdot|\) denotes statistical distance and \(\U_m\) is the uniform random variable.
\end{definition}

In order for extraction to be possible, the most basic necessary assumption is that each source \(\X\in\mathcal{X}\) contains some \emph{randomness}. In this paper, as is standard, we use \emph{min-entropy} as our measure for randomness:
For a source $\X$ with support $\Omega$, we define its min-entropy as $\minH(\X) = -\log(\max_{x\in \Omega}\Pr(\X = x))$.
Note that for any $\X\sim \zo^n$, it holds that $0\le \minH(\X)\le n$.

Unfortunately, randomness alone is not enough to enable extraction. In particular, it is well-known that there do not exist extractors for arbitrary distributions, even when they have a lot of randomness (min-entropy $n-1$). To overcome this, a large body of work has been dedicated to extracting randomness from distributions that not only have some min-entropy, but also exhibit some structure. Two such widely studied classes of structured sources are: (1) \emph{Samplable sources}, i.e., sources that are generated by feeding uniform random bits into low complexity classes such as $\AC^0$ circuits, decision trees, local sources, branching programs, and more \cite{trevisanvadhan00samplable, kamp2011deterministic, deW12local, viola14extractor, chattopadhyayG21space, alrabiah2022low}; and (2) \emph{Recognizable sources}, i.e., sources that are uniform over sets that can be recognized by low complexity classes \cite{shaltiel2011weak,kinne2012pseudorandom,lizuckerman2019variety}.\footnote{A set $S$ is recognized by a class of functions $\cC$ if there exists some $f\in \cC$ such that $S = \{x : f(x) = 0\}$.} These studies have provided important insight into the structure of such complexity classes. Moreover, there is an argument to be made that in nature, most distributions are likely to be generated by such low complexity classes \cite{trevisanvadhan00samplable}. In this paper, we study \emph{algebraic sources}, which are samplable and recognizable sources corresponding to \emph{low degree multivariate polynomials over $\mathbb{F}_2$} (another natural low-complexity class).

\paragraph{Algebraic sources}
Two different flavors of algebraic sources have been studied: variety sources and polynomial sources (corresponding to recognizable and samplable sources, respectively). The task of constructing extractors for these sources has many nice motivations, and can furthermore help us gain important structural insight into low-degree polynomials.

Extractors for polynomial sources (over $\mathbb{F}_2$) with $\poly(\log n)$ degree would immediately yield extractors for sources sampled by $\AC^0[\oplus]$ circuits, based on well-known approximations of such circuits by polynomials  \cite{razborov1987lower, smolensky1993representations}.\footnote{$\AC^0[\oplus]$ circuits are constant depth, polynomial sized circuits with unbounded fan-in AND, OR, NOT, and PARITY gates.} To the best of our knowledge, there are no known nontrivial explicit extractors for sources sampled by such circuits.

Extractors for variety sources (over $\mathbb{F}_2$) have important applications in circuit lower bounds. If one can construct explicit extractors (or even dispersers - \cref{def:disperser}) against degree $2$ varieties with min-entropy $0.01n$, or against degree $n^{0.01}$ varieties with min-entropy $0.99n$, then one immediately obtains new state-of-the-art circuit lower bounds \cite{golovnevk16, golovnevKW21circuit}.

With these motivations in hand, let's proceed to formally define these sources.


\begin{definition}[Polynomial sources]
A degree $d$ polynomial source $\X\sim\F_q^n$ is associated with a polynomial map $P = (p_1, \dots, p_n)$ where each \(p_i:\F_q^m\to\F_q\) is a polynomial of degree at most \(d\). Then, $\X = P(\U_m)$  where $\U_m$ is the uniform distribution over $\F_q^m$.
\end{definition}

\begin{definition}[Variety sources]
A degree $d$ variety source $\X\sim\F_q^n$ is associated with a polynomial map $P = (p_1, \dots, p_m)$ where each \(p_i:\F_q^n\to\F_q\) is a polynomial of degree at most \(d\). Then, $\X$ is uniform over the set of common zeroes of these polynomials $V = \{x\in \F_q^n \textrm{ s.t. } \forall i \in [m]: p_i(x) = 0\}$.
\end{definition}


In this paper, we introduce and study a natural class of sources that is a subclass of polynomial sources and the widely studied NOBF (non-oblivious bit-fixing) sources. Surprisingly, as we will soon see, it is also a subclass of variety sources (\cref{claim:poly-nobf-is-variety-source}).

\begin{definition}[Polynomial NOBF sources]
A degree $d$ polynomial NOBF source $\X\sim\F_q^n$ with min-entropy \(k\in\N\) has the following structure:
\begin{enumerate}
    \item 
    There exists a set $G\subseteq [n]$ with $|G| = k$ that we call the good coordinates of $\X$. These good coordinates in $\X$ are sampled uniformly and independently at random.

    \item 
    Each coordinate outside $G$ is a deterministic function of the $k$ good coordinates of $\X$. Moreover, each such deterministic function is a degree $d$ polynomial.
\end{enumerate}
\end{definition}

We make a basic observation regarding polynomial NOBF sources, which seems to apply to most classes of samplable, NOBF, and recognizable sources.

\begin{claim}
\label{claim:poly-nobf-is-variety-source}
If $\X\sim\F_q^n$ is a degree $d$ polynomial NOBF source, then it is also a degree $d$ polynomial source and a degree $d$ variety source.
\end{claim}

\begin{proof}
Let $\minH(\X) = k$ and the bad positions be specified by polynomials $p_1, \dots, p_{n-k}$. Since the $k$ good positions are degree $1$ polynomials over $x_1, \dots, x_k$ and the $n-k$ bad positions are degree $d$ polynomials over $x_1, \dots, x_k$, $\X$ is indeed a degree $d$ polynomial source.

Now, without loss of generality, assume the first $k$ positions of $\X$ are the good positions and last $n-k$ positions are the bad positions. Consider the following set of polynomial equations over variables $y_1, \dots,y_n$:
\begin{align*}
y_{k+1} - p_1(y_1,\dots, y_k) &= 0\\
&\vdots\\
y_{n} - p_{n-k}(y_1,\dots, y_k) &= 0
\end{align*}
Note $\X$ is uniform over the variety defined by these equations, and is thus also a degree $d$ variety source.
\end{proof}



\paragraph{Related work}
Degree $1$ polynomial and variety sources are known as \emph{affine sources}. Affine sources have been widely studied over both $\F_2$ and larger finite fields $\F_q$ \cite{bourgain2007construction, gabizonR08affine, rao09affine, barakkssw10, devosg10, li11affine, shaltiel11affine, yehudayoff11affine, ben-sassonk12, bourgainDL16affine, li16affine, chattopadhyayGL21, guoVJZ23imageofvariety}. Recently, \cite{li2023optimal} constructed affine extractors over $\F_2$ with asymptotically optimal dependence on min-entropy.

More generally, \cite{dvirGW09polynomial} initiated the study of extractors for polynomial sources. Their extractors work for when $q$ is prime and $q \ge \poly(n, d, k)^{O(k)}$.
\cite{ben-sassonG13polynomial} used sum product estimates and constructed extractors for degree $2$ polynomial sources with min-entropy $cn\log q$ for any constant $c > 0$ and when $q \ge O(1)$.
They also constructed dispersers for arbitrary multilinear polynomials over $\F_4$ with min-entropy $\ge (n/2 + O(1))\log 4$. Extractors for variety sources were first constructed  by \cite{dvir12variety}. They constructed extractors for when either $q \ge d^{\poly(n)}$
or $q\ge \poly(d)$ and min-entropy $\ge (1/2 + \delta)\log(q)$ for any constant $\delta > 0$.
Recently, \cite{guoVJZ23imageofvariety} constructed extractors for images of varieties over $\F_q$ when $q \ge \poly(n, d)$ with no min-entropy restrictions (they define the degree parameter $d$ differently). Over $\F_2$, \cite{remscrim16extractor} constructed extractors for degree $n^{\delta_1}$ varieties with min-entropy $\ge n - n^{\delta_2}$ for arbitrary $\delta_1 + \delta_2 < \frac{1}{2}$.
\cite{cohental15structural} showed that, using optimal affine extractors, one can construct extractors for degree $d$ varieties with min-entropy $n - (n^{1 / (d-1)!} / \log n)^{1/e}$ where $e = \sum_{r=0}^{\infty} \frac{1}{r!} \approx 2.71828$.
Using correlation bounds against low degree polynomials, \cite{dvir12variety, ChattopadhyayT19correlation,  lizuckerman2019variety} constructed extractors for constant degree $d$ variety sources over $\F_2$ with min-entropy $\ge n - \Omega_d(n)$.


We reiterate that before our work, no extractors were constructed for polynomial sources over $\F_2$, even for min-entropy $k\ge n-1$!

\subsection{Our results}

In our main theorem, we construct the first nontrivial extractors for polynomial sources over \(\F_2\).

\begin{theorem}[Explicit extractor for polynomial sources, informal version of \cref{thm:k-wise-independence-poly-time-extractor}]
\label{thm:k-wise-independence-poly-time-extractor-informal}
Let $\eps > 0$ be an arbitrary constant. For all constant $d\in \N$, there exists an explicit $\eps$-extractor $\Ext: \F_2^n\rightarrow \F_2^{\Omega(\log \log n)}$ for degree $d$ polynomial sources over $\F_2$ with min-entropy $k\geq n - \Omega\left(\frac{\sqrt{\log n}}{(\log \log n / d)^{d/2}}\right)$.
\end{theorem}

Prior to our work, there were no known constructions of extractors for polynomial sources over \(\F_2\) that worked for degree \(d>1\) and min-entropy \(k=n-1\). Indeed, all prior constructions required the field size \(q\) to be large, or the degree $d$ to be \(1\).


As polynomial sources can have arbitrarily large input length, it's not clear what is the size of the class of degree $d$ polynomial sources. Therefore, it is unclear if an extractor should even exist for this class. To get around this problem, we come up with an input reduction technique that allows us to bound the number of inputs to the polynomial source by the min-entropy of the source. We view this as our main technical contribution, and it is the key ingredient behind our explicit extractor in \cref{thm:k-wise-independence-poly-time-extractor-informal}.

\begin{lemma}[Input reduction, informal version of \cref{lem:new-input-reduction-lemma}]
\label{cor:new-input-reduction-lemma-informal}
Every degree $d$ polynomial source with min-entropy $k$ (and an arbitrary number of input variables) is $2^{-k}$-close to a convex combination of polynomial sources with min-entropy $k-1$ and $O(k)$ input variables.
\end{lemma}
Recall that $\X$ is a convex combination of distributions $\{\Y_i\}$ if there exist probabilities $\{p_i\}$ summing up to $1$ such that $\X = \sum_i p_i\Y_i$.
It is well known that an $\eps$-extractor for $\{\Y_i\}$ will also be an $\eps$-extractor for $\X$.
Hence, this lemma reduces the task of extracting from polynomial sources with an arbitrary number of input variables to the task of extracting from polynomial sources with $O(k)$ input variables.

We also show negative results for polynomial NOBF sources against sumset extractors. Sumset extractors are extremely powerful and can be used to extract not only from sumset sources but also, using reductions to sumset sources, from many well studied models of weak sources such as degree $1$ polynomial / variety sources (affine sources), two independent sources, sources generated by branching programs, sources generated by $\AC^0$ circuits, and many more \cite{chattopadhyayL22sumset}. Let's first define sumset sources:
\begin{definition}
  A source $\X$ is a $(k, k)$ sumset source if $\X = \A + \B$, where  $\A, \B$ are independent distributions over $\zo^n$ with $\minH(\A) \ge k, \minH(\B)\ge k$, and $+$ denotes bitwise xor.
\end{definition}
Note that $k\le \minH(\X) \le 2k$ and so, $\minH(\X) = \Theta(k)$. When we write $X$ is a sumset source of min-entropy $k$, we actually mean $\X = \A + \B$ where $\minH(\A)\ge k, \minH(\B)\ge k$.
Recently, sumset extractors with an asymptotically optimal dependence on min-entropy  $\left(O(\log n)\right)$ were constructed by Li \cite{li2023optimal}. A natural question is whether various algebraic sources can be reduced to sumset sources. We show here that sumset extractors cannot even \emph{disperse} (let alone extract) from quadratic NOBF sources with very high min-entropy.

\begin{theorem}[Sumset extractor lower bound, informal version of 
\cref{thm:quadratic-nobf-sumset-extractor-lower-bound}]
\label{thm:quadratic-nobf-sumset-extractor-lower-bound-informal}
Sumset extractors cannot be used to disperse from degree $2$ polynomial NOBF sources with min-entropy $n - O\left(\frac{n}{\log\log n}\right)$.
\end{theorem}
As polynomial NOBF sources are both variety \emph{and} polynomial sources, this also implies that sumset extractors cannot be used to extract from degree $2$ variety sources or degree $2$ polynomial sources over $\F_2$ with min-entropy $n - O\left(\frac{n}{\log\log n}\right)$. For degree $2$ variety sources over $\F_2$, one can use a sumset extractor to extractor above min-entropy $n - \Omega\left(\frac{n}{\log n}\right)^{1/e}$ \cite{cohental15structural}. On  the other hand, using \emph{correlation bounds}, one can construct explicit extractors against degree $2$ varieties with min-entropy $(1-c)n$ for some small constant $c>0$ \cite{dvir12variety, ChattopadhyayT19correlation, lizuckerman2019variety}. Thus, the above result shows that sumset extractors cannot be used to get better extractors than what we get using correlation bounds against low degree polynomials. We find this surprising as it implies that the generalized inner product function is a better extractor for degree $2$ variety sources than any optimal (blackbox) sumset extractor.


\paragraph{Organization}
The rest of the paper is organized as follows. In \cref{sec:proof-overview}, we give an overview of our proofs. In \cref{sec:prelims}, we provide basic definitions and useful properties that we will use later.
In \cref{sec:construct-extractor}, we prove our input reduction lemma (\cref{cor:new-input-reduction-lemma-informal}), and use it to obtain our explicit extractor for polynomial sources (\cref{thm:k-wise-independence-poly-time-extractor-informal}). In \cref{sec:impossibility}, we prove our result on the limitations of sumset extractors for extracting from quadratic NOBF sources (\cref{thm:quadratic-nobf-sumset-extractor-lower-bound-informal}). In \cref{sec:open}, we conclude with various open problems.

\section{Overview of our techniques}
\label{sec:proof-overview}

In this section, we sketch the proofs of all our main results.

\subsection{Existential results}

To warm up, it is not clear whether a random function is a good extractor for degree $d$ polynomial sources. Usually, such proofs proceed by arguing that for a fixed source of min-entropy $k$, a random function is an $\varepsilon$-extractor with probability at least $1 - 2^{-2^{k}\varepsilon^2}$. Then, one can do a union bound over the total number of sources in the class to obtain that a random function is a good extractor. The main issue that arises for the class of polynomial sources is that the number of input variables to the polynomials can be arbitrary. Thus, the exact size of this class is not clear. To overcome this difficulty, we use our input reduction lemma (\cref{cor:new-input-reduction-lemma-informal}).
Using this, it suffices to consider degree $d$ polynomial sources with $O(k)$ inputs. This class of polynomial sources has size $2^{O(k)^d\cdot n}$. Thus, the  the earlier union bound-based argument now works out:

\begin{seclemma}[Informal version of \cref{lem:image-extractor-exists}]
\label{lem:image-extractor-exists-informal}
A random function is a $2^{-\Omega(k)}$-extractor for the class of degree $d$ polynomial sources over $\F_2$ with min-entropy $k\ge O(d + \log n)$.
\end{seclemma}

\subsection{Input reduction}

We will now sketch the proof for the input reduction lemma that was used in the existential result, above. (In fact, it will also be crucially used in our explicit construction.) We begin by showing that for any polynomial map $f(\U_m)$, there exists a full rank linear function $L$ and a fixing $b$ of $L$ such that $f(\U_m) \approx_{\eps} f(\U_m) | L(\U_m) = b$. In fact, we show the stronger claim that most such fixings $b$ work:


\begin{seclemma}[Existence of affine white-box PRGs]\label{lem:new-version-affine-PRG-informal}
For any polynomial map \(f:\F_2^m\to\F_2^n\) and \(0<\eps<1/4\), there exists a full rank linear function \(L:\F_2^m\to\F_2^{m-\ell}\) with \(\ell=n+3\log(1/\eps)\) such that
\[
|f(\U_m)\circ L(\U_m)-f(\U_m)\circ\U_{m-\ell}|\leq 2\eps,
\]
where \(\U_m\) and \(\U_{m-\ell}\) are independent.
\end{seclemma}

\begin{proof}[Proof sketch]
We show a random \(L\) works. Indeed, by definition of statistical distance, note that
\[
|f(\U_m)\circ L(\U_m)-f(\U_m)\circ\U_{m-\ell}|
= \E_{z\sim f(\U_m)}\left[|(L(\U_m)\mid f(\U_m)=z) - \U_{m - \ell}|\right].
\]

We now apply the min-entropy chain rule (\cref{lem:min-entropy-chain-rule}) to infer that with high probability over fixings of $f$ to $z$, the input distribution to $L$ will have high min-entropy. Indeed, there will exist some distribution $\X$ with min-entropy at least \(k = m - n - \log(1/\eps) = m - \ell + 2\log(1/\eps)\) such that 
\[
 \E_{z\sim f(\U_m)}\left[|(L(\U_m)\mid f(\U_m)=z) - \U_{m - \ell}|\right] \le \eps + |L(\X) - \U_{m - \ell}|\\
\]
As $L$ was initially chosen as a random function, we apply the leftover hash lemma (see \cref{cor:nice-corollary-leftover-hash-lemma}) to infer that $|L(\X) - \U_{m - \ell}|\le \eps$ as desired. Furthermore, this implies that $L$ will have full rank because otherwise, $|L(\U_m) - \U_{m-\ell}| \ge \frac{1}{2}$, which contradicts the fact that \(2\eps<1/2\).
\end{proof}

Note that this lemma already yields an input reduction to a single polynomial source with $O(n)$ variables. This can be done by fixing the output of $L$ to some $b$ such that $|(f(\U_m) \mid L(\U_m) = b) - f(\U_m)| \le 2\eps$. Once we fix the output of $L$, we induce $m - \ell$ affine constraints on the input variables. As polynomial sources are closed under affine restrictions, the resulting polynomial map is still a degree $d$ polynomial map and the resulting distribution is still close enough to the original one, as desired. However, we can do better.

We will first prove the following helpful claim. This claim shows there exists a way to map every source \(\X\sim\zo^n\) with min-entropy $k$ source to a source over $k+1$ bits without decreasing the min-entropy.

\begin{secclaim}\label{cl:entropy-smoothing-informal}
Let \(\X\sim\zo^n\) be a polynomial source with min-entropy at least \(k>0\). Then there exists a function \(S:\zo^n\to\zo^{k+1}\) such that \(S(\X)\) has min-entropy at least \(k\).
\end{secclaim}

This claim is actually true for arbitrary sources $\X$ and we prove it by a simple case analysis on probabilities of the smallest two elements in support of $\X$ (see \cref{cl:entropy-smoothing} for further details). Using this claim, we are ready to sketch the proof of our main lemma that will help us achieve the input reduction:

\begin{seclemma}\label{lem:new-version-affine-PEG-informal}
For any polynomial source \(f:\F_2^m\to\F_2^n\) where \(f(\U_m)\) has min-entropy at least \(k\), there exists a linear function \(L:\F_2^m\to\F_2^{m-O(k)}\) such that
\[
\Pr_{b\sim L(\U_m)}\biggl[H_\infty(f(\U_m)\mid L(\U_m)=b) \geq k-1\biggl] \geq 1-2^{-k}.
\]
\end{seclemma}

\begin{proof}[Proof sketch]
Let \(S:\F_2^n\to\F_2^{k+1}\) be a function guaranteed to exist by above claim so that \(S(f(\U_m))\) has min-entropy at least \(k\).
Using data processing inequality (see \cref{cl:data-processing-entropy}), it suffices to show that \(S(f(\U_m))\) has high enough min-entropy with high probability over fixing \(L(\U_m)\).
Let $\Y = \U_m$.
By \cref{lem:new-version-affine-PRG-informal}, there exists \(L:\F_2^m\to\F_2^{m-\ell}\) with \(\ell=k+4+3\log(1/\eps)\) such that:
\[
|S(f(\Y))\circ L(\Y) - S(f(\Y))\circ\U_{m-\ell}|\leq \eps
\]
This implies 
\[
\E_{b\sim L(\U_m)}\left[|(S(f(\Y))\mid L(\Y) = b) - S(f(\U_m))|\right] \le \eps
\]
By Markov's inequality, we infer that 
\[
\Pr_{b\sim L(\U_m)}\left[|(S(f(\Y))\mid L(\Y) = b) - S(f(\U_m))|\ge \sqrt{\eps}\right] \le \sqrt{\eps}
\]

Setting \(\eps=2^{-2k}\), every element in support of the distribution \(S(f(\Y\mid L(\Y)=b))\) must occur with probability at most \(2^{-k}+\sqrt{\eps}\leq 2^{-k+1}\), and thus has min-entropy at least \(k-1\). The result follows.
\end{proof}

Using this, we are ready to prove our input reduction lemma:

\begin{proof}[Proof of \cref{cor:new-input-reduction-lemma-informal}]
Using \cref{lem:new-version-affine-PEG-informal}, most fixings of $L$ leave $f$ with min-entropy at least $k-1$. These good fixings form a convex combination of such sources $f$. As argued earlier, such $L$ induces $m - O(k)$ linear fixings on the input variables and hence the resulting polynomial map in each of these convex combinations is over $O(k)$ variables and has degree $d$, as desired.
\end{proof}

\subsection{Explicit construction}

We sketch here the proof of a slightly weaker result that illustrates our main idea.
\begin{sectheorem}[Weaker version of \cref{thm:k-wise-independence-poly-time-extractor-informal}]
For all constant degree $d\in \N$, there exists an explicit extractor $\Ext:\F_2^n\rightarrow \F_2$ for polynomial sources with min-entropy $k \ge n - \Omega(\log\log n)$.
\end{sectheorem}

\begin{proof}[Proof sketch]
Let \(\X\) be a degree \(d\) polynomial source with \(m\) inputs, \(n\) outputs, and min-entropy $n - g$ where $g = O(\log \log n)$. Consider a small length $t = 2g$ prefix of the output bits, and let this source be $\X_{pre}$. We observe that $\X_{pre}$ has min-entropy at least $t-g = t/2$ (see \cref{prop:project-source-min-entropy}). We now use our input reduction lemma (\cref{cor:new-input-reduction-lemma-informal}) over $\X_{pre}$ to infer that it is close to a convex combination of degree $d$ polynomial sources with $O(t)$ inputs and min-entropy $t/2-1$.
Hence, it suffices to construct an extractor for min-entropy $t/2 - 1$ degree $d$ polynomial sources with $O(t)$ inputs and $t$ outputs.

By our existential results (\cref{lem:image-extractor-exists-informal}), a random function over $t$ bits will be an extractor for such sources. We exhaustively try all the $2^{2^t}$ functions from $t$ bits to $1$ bits as our candidate extractor. We brute force search over all the $2^{O(t)^d\cdot t}$ degree $d$ polynomial sources with $O(t)$ inputs and $t$ outputs. Then, for each of them, we check if it has enough min-entropy. If it does, we input the source into our candidate extractor and check if the output is close to uniform. We will eventually find a candidate extractor that will work for all such sources, and we output that function as our extractor.

The time required by the above procedure is $2^{2^t + O(t)}$. As $t = O(\log \log n)$, the above procedure indeed runs in $\poly(n)$ time. 
\end{proof}

In our actual construction, we achieve better parameters by brute forcing over all $r$-wise independent functions (for very large $r$) as our candidate extractor instead of all functions. We again take advantage of the fact that the input reduction lemma actually reduces number of input variables to $O(k)$, making the class of polynomial sources that we have to brute force over even smaller.
Together, these optimizations allow us to handle smaller min-entropy. See \cref{thm:k-wise-independence-poly-time-extractor} for further details.

\subsection{Impossibility results}

All our impossibility results are against polynomial NOBF sources and hence apply (via \cref{claim:poly-nobf-is-variety-source}) to both polynomial sources and variety sources. We show that sumset extractors, arguably the most powerful general purpose extractors, cannot be used to even disperse from degree $2$ polynomial NOBF sources below min-entropy $n - O\left(\frac{n}{\log \log n}\right)$ (\cref{thm:quadratic-nobf-sumset-extractor-lower-bound-informal}). These results are formally proven in \cref{subsec:nobf-lower-bound-sumset-disperser} and \cref{subsec:nobf-lower-bound-sumset-extractor}.
We will use the following useful theorem to show this. This theorem states there exists some quadratic NOBF source which does not contain any sumset source of small min-entropy.

\begin{sectheorem}[Informal version of \cref{thm:quadratic-nobf-sumset-disperser-lower-bound}]
There exists a degree $2$ polynomial NOBF source $\X\sim\F_2^n$ with $\minH(\X) = n - O\left(\frac{n}{\log\log n}\right)$ such that for all $\A, \B\sim\F_2^n, \minH(\A) \ge \Omega(\log\log n), \minH(\B) \ge \Omega(\log\log n)$, it holds that $\supp(\A) + \supp(\B) \not\subset \supp(\X)$.
\end{sectheorem}

\begin{proof}[Proof sketch]
We take the $n-k$ bad bits in $\X$ to be random degree $2$ polynomials.
Say such $\A, \B$ exist and let $C, D$ be projections of $\supp(\A), \supp(\B)$ respectively onto the good bits of $\X$. Let $P:\F_2^k\rightarrow \F_2^{n-k}$ be the polynomial map of the bad bits. Then, it holds that $P(C) + P(D) = P(C+D) + y$ for some $y\in \F_2^{n-k}$. To simplify presentation, assume for this proof sketch that $y = 0^{n-k}$.
We first observe the following: 
\begin{secclaim}[Informal version of \cref{prop:poly-map-sumset-is-affine}]
There exist affine subspaces $U, V$ such that $P(U) + P(V) = P(U+V)$ and $|U| \ge |C|, |V| \ge |D|$.
\end{secclaim}
Hence, without loss of generality, we can assume that $C$ and $D$ are affine subspaces. We now use a probabilistic argument to show there exists a quadratic map where such large affine subspaces $C$ and $D$ cannot exist.


\begin{secclaim}[Informal version of \cref{prop:random-poly-map-no-affine-sumset}]
There exists a degree $2$ polynomial map $P:\F_2^k\rightarrow \F_2^{n-k}$ such that for every pair of affine subspaces $U, V$, both of dimensions $ \ge \Omega(\log\log n)$, there exist $u\in U, v\in V$ such that $P(u) + P(v) \ne P(u+v)$.
\end{secclaim}
Hence, the sumset property is violated and we get a contradiction.
\end{proof}

Using these, we finally present the proof of our lower bound result:

\begin{proof}[Proof sketch of \cref{thm:quadratic-nobf-sumset-extractor-lower-bound-informal}]
Let $\X$ be the degree $2$ polynomial NOBF source with min-entropy $n - O\left(\frac{n}{\log\log n}\right)$ that doesn't contain any sumset of min-entropy $O(\log \log n)$. We apply a bipartite Ramsey bound (\cref{lem:bipartite-ramsey-bound}), to show that if a quadratic NOBF source doesn't contain sumsets where each of the two sets has size $s$, then it has small intersection with sumsets where each of the two sets has size $O(2^s)$ (see \cref{lem:worst-case-average-case-sumset} for details). Setting $s = O(\log \log n)$, we infer that $\X$ has very small intersection with sumset sources of min-entropy $\Omega(\log n)$. From this, we infer $\X$ is far away from any convex combination (see \cref{def:convex-combination}) of sumset sources with min-entropy $\Omega(\log n)$. As sumset extractors below min-entropy $O(\log n)$ cannot exist (every function is constant on $\Omega(\log n)$ dimensional affine subspace), this shows we cannot use sumset extractors to even disperse against quadratic NOBF sources. See \cref{thm:quadratic-nobf-sumset-extractor-lower-bound} for further details.
\end{proof}

\dobib

%% file: sections/prelims.tex
\section{Preliminaries}
\label{sec:prelims}

To simplify notation, we will use $\circ$ to mean concatenation. Also, all logs in this paper are base $2$.

\subsection{Basic probability lemmas}

Given two random variables \(\X,\Y\), we let \(|\X-\Y|\) denote their statistical distance, defined as
\[
|\X-\Y|:=\max_S[\Pr[\X\in S]-\Pr[\Y\in S]]=\frac{1}{2}\sum_{z}\left|\Pr[\X=z]-\Pr[\Y=z]\right|.
\]

We write \(\X\approx_\eps\Y\) and say that \(\X,\Y\) are \emph{\(\eps\)-close} if \(|\X-\Y|\leq\eps\), and we write \(\X\equiv\Y\) if \(|\X-\Y|=0\).

We will often use the fact that applying a function can only decrease the distance between two distributions:
\begin{secfact}[Data-processing inequality]
\label{fact:data-processing-inequality}
For any random variables \(\X,\X^\pr\sim X\) and function \(f:X\to Y\),
\[
|\X-\X^\pr|\geq|f(\X)-f(\X^\pr)|.
\]    
\end{secfact}

We will utilize the well known fact that for any two distributions $\X, \Y$, with high probability, fixings of $\Y$ decrease min entropy of $\X$ by about $\log(|\supp(\Y)|)$:
\begin{seclemma}[Min-entropy chain rule]\label{lem:min-entropy-chain-rule}
    For any random variables \(\X\sim X\) and \(\Y\sim Y\) and \(\eps>0\),
    \[
    \Pr_{y\sim\Y}[H_\infty(\X\mid\Y=y)\geq H_\infty(\X)-\log|\supp(\Y)|-\log(1/\eps)]\geq1-\eps.
    \]
\end{seclemma}

\subsection{Extractors}
We start by defining \emph{dispersers}, which are a weaker version of extractors. While the output of an extractor must look nearly uniform, the output of a disperser only needs to be \emph{non-constant}.

\begin{secdefinition}[Disperser]
\label{def:disperser}
A function $\Disp:\zo^n\rightarrow \zo$ is a disperser for a class of distributions $\cX$ if for all $\X\in \cX$, the set $\{\Disp(\X)\} = \zo$.
\end{secdefinition}

While the main purpose of this paper is to construct seedless extractors (for polynomial sources), it turns out that \emph{seeded} extractors will also be useful in our arguments. We define them, below.

\begin{secdefinition}[Seeded extractor]\label{def:strong-seeded-extractor}
We say that a deterministic function \(\sExt:\zo^m\times\zo^s\to\zo^r\) is a \emph{\((k,\eps)\)-strong seeded extractor} if for any \(\X\sim\zo^m\) with min-entropy at least \(k\),
\[
\sExt(\X,\Y)\circ\Y\approx_\eps\U_r\circ\Y,
\]
where \(\Y\sim\zo^s\) and \(\U_r\sim\zo^r\) are independent uniform random variables. We say \(\sExt\) is \emph{linear} if the function \(\sExt(\cdot,y):\zo^m\to\zo^r\) is a degree $1$ polynomial, for all \(y\in\zo^s\).
\end{secdefinition}

One classic way to construct seeded extractors is via the following theorem.

\begin{sectheorem}[Leftover Hash Lemma \cite{hill1999pseudorandom}]\label{lem:leftover-hash-lemma}
    Let \(\mathcal{H}=\{H:\zo^n\to\zo^m\}\) be a \(2\)-universal hash family with output length \(m=k-2\log(1/\eps)\), meaning that for any \(x\neq y\), \(\Pr_{H\sim\mathcal{H}}[H(x)=H(y)]\leq2^{-m}.\) Then the function \(\sExt:\zo^n\times\mathcal{H}\to\zo^m\) defined as \[\sExt(x,h):=h(x)\] is a \((k,\eps)\)-strong seeded extractor.
\end{sectheorem}
\begin{seccorollary}\label{cor:nice-corollary-leftover-hash-lemma}
For any \(\eps>0\) and \(m=k-2\log(1/\eps)\), the function \(\sExt:\zo^n\times\F_2^{m\times n}\to\zo^m\) defined as
    \[
    \sExt(x,L):=Lx
    \]
    is a linear \((k,\eps)\)-strong seeded extractor.
\end{seccorollary}
\begin{proof}
    It suffices to show that the family of all linear functions \(L:\F_2^n\to\F_2^m\), which correspond to matrices \(\F_2^{m\times n}\), is a \(2\)-universal hash family. That is, we must show that for any distinct \(x,y\),
    \[
    \Pr_{L\sim\F_2^{m\times n}}[L(x)=L(y)]=\Pr_L[L(x+y)=0]\leq2^{-m}.
    \]
    This is equivalent to showing that \(\Pr_L[Lx=0]\leq2^{-m}\) for any nonzero \(x\). This is clearly true (in fact, equality holds), since the rows of \(L\) are exactly \(m\) independent uniform parity checks on a nonzero \(x\).
\end{proof}


Next, we define the notion of \emph{reductions} for extractors.

\begin{secdefinition}[Convex combination]
\label{def:convex-combination}
We say $\X$ is a convex combination of distributions $\{\Y_i\}$ if there exist probabilities $\{p_i\}$ summing up to $1$ such that $\X = \sum_i p_i\Y_i$.
\end{secdefinition}

\begin{secfact}
\label{fact:convex-combination}
Let $\Ext: \zo^n\rightarrow \zo$ be an $\varepsilon$-extractor for a class of distributions $\cX$. Let $\X\sim \zo^n$ be a distribution that can be written as convex combination of distributions in $\cX$. Then, $\Ext$ is also an $\varepsilon$-extractor for $\X$.
\end{secfact}

Finally, the following proposition shows that for a fixed source, a random function is a good extractor.

\begin{secproposition}[Implicit in \protect{\cite[Theorem 2.5.1]{rao2007thesis}}]
\label{prop:random-function-good-extractor}
For every $n, m\in \N$, every $k \in [0, n]$, every $\eps > 0$, and every $\X\sim \zo^n$ with $\minH(\X) = k$, if we choose a random function $\Ext: \zo^n\rightarrow \zo^m$ with $m \le k - 2\log(1 / \eps) - O(1)$, then $\Ext(\X) \approx_{\eps} \U_m$ with probability $1 - 2^{-\Omega(K\eps^2)}$ where $K = 2^k$.
\end{secproposition}

\dobib

%% file: sections/constructions.tex
\section{Constructing extractors}
\label{sec:construct-extractor}

In this section we construct extractors for polynomial sources, both existentially and explicitly.

\subsection{Input reduction}

As mentioned earlier, the key ingredient behind our constructions is an \emph{input reduction lemma}, which allows us to assume that the number of inputs to a polynomial source is small. More formally, we prove the following, which shows that any polynomial source is close to a convex combination of polynomial sources which have roughly the same min-entropy, but a much shorter input length.
\begin{sectheorem}[Input reduction]\label{lem:new-input-reduction-lemma}
Let \(\X\sim\F_2^n\) be a degree \(d\) polynomial source with min-entropy at least \(k\). Then \(\X\) is \(2^{-k}\)-close to a convex combination of degree \(d\) polynomial sources with min-entropy at least \(k-1\) and input length at most \(11k\).
\end{sectheorem}

It is well-known that any extractor that works for a family of sources \(\mathcal{X}\) also works for a convex combination of sources from $\cX$. Thus, the above input reduction lemma allows us to just focus on constructing extractors for polynomial sources with input length that is linear in the min-entropy of the source. Now, in order to prove \cref{lem:new-input-reduction-lemma}, we begin with the following useful lemma.


\begin{seclemma}[Existence of affine white-box PRGs]\label{lem:new-version-affine-PRG}
For any function \(f:\F_2^m\to\F_2^n\) and \(0 < \eps < \frac{1}{4}\), there exists a full rank linear function \(L:\F_2^m\to\F_2^{m-\ell}\) with \(\ell=n+3\log(1/\eps)\) such that
\[
|f(\U_m)\circ L(\U_m)-f(\U_m)\circ\U_{m-\ell}|\leq 2\eps,
\]
where \(\U_m\) and \(\U_{m-\ell}\) are independent uniform distributions.
\end{seclemma}


\begin{proof}
We show a random choice of \(L\) works. Indeed, given an independent uniformly random \(L\), we obtain:
\begin{align*}
|f(\U_m)\circ L(\U_m)-f(\U_m)\circ\U_{m-\ell}|
& = \E_{z\sim f(\U_m)}\left[|(L(\U_m)\mid f(\U_m)=z) - \U_{m - \ell}|\right]\\
&\leq\eps + |L(\X) - \U_{m - \ell}|\\
&\leq 2\eps.
\end{align*}
Above, \(\X\) has min-entropy at least \(k = m - n - \log(1/\eps) = m - \ell + 2\log(1/\eps)\) by the min-entropy chain rule (\cref{lem:min-entropy-chain-rule}), and the last inequality follows by the leftover hash lemma (\cref{cor:nice-corollary-leftover-hash-lemma}).
We claim that such a linear function $L$ has full rank. Indeed if not, then $|L(\U_m) - \U_{m - \ell}| \ge \frac{1}{2} > 2\eps$, a contradiction.
\end{proof}

The above lemma can easily be used to show that the input length of a polynomial source can be reduced to \(O(n)\). To prove the more challenging result that the input length can be reduced to \(O(k)\) (\cref{lem:new-input-reduction-lemma}), we need a few more tools, starting with the following claim.

\begin{secclaim}[Entropy smoothing]\label{cl:entropy-smoothing}
For every random variable \(\X\sim\zo^n\) with min-entropy at least \(k\), there exists a function \(S:\zo^n\to\zo^{k+1}\) such that \(S(\X)\) has min-entropy at least \(k\).
\end{secclaim}
\begin{proof}
    We may assume that \(k>0\), because otherwise any constant function \(S:\zo^n\to\zo\) works. Now, consider the two least probable elements \(x_1,x_2\in\supp(\X)\) that occur with probabilities \(0<p_1\leq p_2<1\), respectively. We consider two cases.
    \begin{casesenum}
        \item \(p_1+p_2>2^{-k}\). In this case, we know that each \(x\in\supp(\X)\setminus\{x_1\}\) occurs with probability \(>2^{-k-1}\), and thus \(|\supp(\X)\setminus\{x_1\}|<2^{k+1}\). As a result, we have \(|\supp(\X)|\leq2^{k+1}\), and we are done.
        \item \(p_1+p_2\leq2^{-k}\). In this case, we merge \(x_1,x_2\) into a single support element that gets hit with probability at most \(2^{-k}\). This yields a new random variable that still has min-entropy at least \(k\), but whose support has one less element than \(\supp(\X)\). At this point, the support either has size at most \(2^{k+1}\) (and we are done), or we can recurse on this same argument until we eventually hit the first case (which is guaranteed to eventually happen, since the support is shrinking at each step).\qedhere
    \end{casesenum}
\end{proof}

We will also use the data processing inequality for min-entropy. We provide its proof, for completeness. 

\begin{secclaim}[\protect{\cite[Lemma 6.8]{vadhan12pseudorandomness}}]\label{cl:data-processing-entropy}
For any random variable \(\X\) and function \(f\),
\[
H_\infty(f(\X))\leq H_\infty(\X).
\]
\end{secclaim}
 \begin{proof}
 For any given \(x\), we of course have \(\Pr[f(\X)=f(x)]\geq\Pr[\X=x]\). The result follows immediately by considering the element \(x\in\supp(\X)\) that is hit with the highest probability.
 \end{proof}

Equipped with these, we prove our main lemma:


\begin{seclemma}[Existence of affine white-box PEGs]\label{lem:new-version-affine-PEG}
For any function \(f:\F_2^m\to\F_2^n\) such that \(f(\U_m)\) has min-entropy at least \(k > 1\), there exists a full rank linear function \(L:\F_2^m\to\F_2^{m-11k}\) such that
\[
\Pr_{b\sim L(\U_m)}\biggl[H_\infty(f(\U_m)\mid L(\U_m)=b) \geq k-1\biggl] \geq 1-2^{-k}.
\]
\end{seclemma}
\begin{proof}
If $m \le 11k$, then we are trivially done. Otherwise, let \(S:\F_2^n\to\F_2^{k+1}\) be a function, guaranteed to exist by \cref{cl:entropy-smoothing}, such that \(S(f(\U_m))\) has min-entropy at least \(k\). By \cref{cl:data-processing-entropy}, it suffices to show that \(S(f(\U_m))\sim\zo^{k+1}\) has high enough min-entropy with high probability over fixing \(L(\U_m)\).
Let $\Y = \U_m$.
By \cref{lem:new-version-affine-PRG}, there exists a full rank function \(L:\F_2^m\to\F_2^{m-\ell}\) with \(\ell=k+4+3\log(1/\eps)\) such that:
\[
|S(f(\Y))\circ L(\Y) - S(f(\Y))\circ\U_{m-\ell}|\leq \eps
\]
This implies 
\[
\E_{b\sim L(\Y)}\left[|(S(f(\Y))\mid L(\Y) = b) - S(f(\U_m))|\right] \le \eps
\]
By Markov's inequality, we infer that 
\[
\Pr_{b\sim L(\Y)}\left[|(S(f(\Y))\mid L(\Y) = b) - S(f(\U_m))|\ge \sqrt{\eps}\right] \le \sqrt{\eps}
\]

Setting \(\eps=2^{-2k}\), we infer that the for every `good' fixing $b$, and $z\in \zo^{k+1}$, it holds that the probability \(S(f(\Y\mid L(\Y)=b))\) outputs $z$ is at most \(2^{-k}+\sqrt{\eps}\leq 2^{-k+1}\). Hence, \(S(f(\Y\mid L(\Y)=b))\) has min-entropy at least \(k-1\). The result follows, since \(\ell=k+4+3\log(1/\eps)=k+4+6k\leq 11k\).
\end{proof}

Using this main lemma, our theorem easily follows:


\begin{proof}[Proof of \cref{lem:new-input-reduction-lemma}]
Let $P: \F_2^m\rightarrow \F_2^n$ be the degree $d$ polynomial map such that $\X = P(\U_m)$.
If $m \le 11k$, then we are done.
Otherwise, we apply \cref{lem:new-version-affine-PEG} to $P$ and infer that there exists a full rank linear function $L: \F_2^m \rightarrow \F_2^{m-11k}$ such that 
\begin{align}\label{eq:fancy-entropy-inequality}
\Pr_{b\sim \F_2^{m-11k}}\biggl[H_\infty(P(\U_m)\mid L(\U_m)=b) \geq k-1\biggl] \geq 1-2^{-k}.
\end{align}
For any $b\in \F_2^{m-11k}$, we can find a degree $d$ polynomial map $Q_b:\F_2^{11k}\rightarrow \F_2^n$ such that $Q_b(\U_{11k}) = (P(\U_m)\mid L(\U_m)=b)$.
Let $G = \{b\in \F_2^{m-11k}: \minH(Q_b(\U_{11k})) \ge k-1\}$. By \cref{eq:fancy-entropy-inequality}, we know that $|G|2^{-(m-11k)} \ge 1 - 2^{-k}$.
We see that:
\[
    P(\U_m) = \sum_{b\in \F_2^{m-11k}} 2^{-(m-11k)} Q_b(\U_{11k}) = \sum_{b\in G} 2^{-(m-11k)} Q_b(\U_{11k}) + (1 - |G|2^{-(m-11k)})\Y
\]
where $\Y = \sum_{b\in \F_2^{m-11k}\setminus G} 2^{-(m-11k)} Q_b(\U_{11k})$.
The claim now follows.
\end{proof}

\subsection{Existential results}

We first show that with high probability, a random function is a good extractor. We will then improve upon it to show that for large enough $t$, a function sampled using a $t$-wise independent distribution is a good enough extractor.

\begin{seclemma}\label{lem:image-extractor-exists}
Let $n, d, k, \varepsilon$ be such that $k \ge O(d + \log(n/\eps))$, and $m = k - 2\log(1/\varepsilon) - O(1)$. Then, there exists a function $\Ext:\F_2^n\rightarrow \F_2^m$ that is an $\eps$-extractor for the class of degree $d$ polynomial sources over $\F_2^n$ with min-entropy at least $k$.
\end{seclemma}





\begin{proof}
By \cref{lem:new-input-reduction-lemma} and \cref{fact:convex-combination}, it suffices to extract from degree $d$ polynomial sources $\X^\pr\sim \F_2^n$ with $O(k)$ inputs and $\minH(\X) \ge k-1$. By \cref{fact:data-processing-inequality}, we infer that an extractor with error $\varepsilon$ for $\X^\pr$ is also an extractor for $\X$ with error $\varepsilon + 2^{-k}$.

By \cref{prop:random-function-good-extractor}, for a fixed source $\Y$ with $\minH(\Y) = k$, a random function $r:\F_2^n\rightarrow \F_2^m$ satisfies $r(\Y) \approx_{\varepsilon} \U_m$ with probability $1-2^{-\Omega(2^k)\varepsilon^2}$ where $m = k - 2\log(1/\varepsilon) - O(1)$. We now do a union bound over all the $2^{\binom{\ell}{\le d}\cdot n}$ degree $d$ sources with $\ell$ inputs and $n$ outputs. As $\varepsilon \ge 2^{-\Omega(k)}, k \ge O(\log n), k \ge O(d), \ell = O(k)$, the union bound indeed succeeds and we infer the claim.
\end{proof}

We will use $t$-wise independent hash functions to help construct extractors for polynomial sources. We will show a random function from a family of such functions will be an extractor. Let us first define them:

\begin{secdefinition}[\protect{\cite[Definition 3.3.1]{vadhan12pseudorandomness}}]
For any $n, m, t\in \N$ such that $t\le 2^n$, we say that a family of functions $\cH = \{h:\zo^n\rightarrow\zo^m\}$ is \emph{$t$-wise independent} if for all fixed distinct $x_1, \dots, x_t\in \zo^n$, it holds that the random variables $h(x_1), \dots, h(x_t)$ are independently and uniformly distributed in $\zo^m$ when $h$ is a randomly chosen function from $\cH$.
\end{secdefinition}

We will rely on the following property of $t$-wise independent hash functions in our construction:
\begin{seclemma}[Implicit in \protect{\cite[Proposition A.1]{trevisanvadhan00samplable}}]\label{lem:t-wise-independent-general-family-extractor}
Let $\cX$ be an arbitrary class of distributions over $n$ bits that have min-entropy at least $k$. Let $\cH$ be a class of $t$-wise independent hash functions from $n$ bits to $m$ bits, where $t = 2\log(k + |\cX|)$ and $m = k - 2\log(1/\varepsilon) - \log(t) - 2$. Then, there exists some $h\in \cH$ such that $h$ is a $(k, \varepsilon)$ extractor against all sources in the class $\cX$.
\end{seclemma}

Using this, we extend our existential result for $t$-wise independent hash functions.
\begin{seccorollary}\label{lem:t-wise-independent-image-extractor-exists}
Let $n, d, k, t, \varepsilon$ be such that $t = 2\log\left(k + 2^{\binom{O(k)}{d}\cdot n}\right), m = k - \log(t) - 2\log(1/\varepsilon) - O(1)$.
Then, there exists a function $\Ext:\F_2^n\rightarrow \F_2^m$ from a family of $t$-wise independent functions that is an $\eps$-extractor for the class of degree $d$ polynomial sources over $\F_2^n$ with min-entropy at least $k$.
\end{seccorollary}


\begin{proof}
As earlier, by \cref{lem:new-input-reduction-lemma} and \cref{fact:convex-combination}, it suffices to extract from degree $d$ polynomial sources $\X^\pr\sim \F_2^n$ with $O(k)$ inputs and $\minH(\X) \ge k-1$. By \cref{fact:data-processing-inequality}, we infer that an extractor with error $\varepsilon$ for $\X^\pr$ is also an extractor for $\X$ with error $\varepsilon + 2^{-k} \le 2\eps$. Using naive bounds on the number of such polynomial sources, there are at most $2^{\binom{O(k)}{\le d}\cdot n}$ such sources.  We apply \cref{lem:t-wise-independent-general-family-extractor} for our choice of parameters and infer the claim.
\end{proof}

\subsection{Explicit construction}

We use the input reduction trick and the existential results to construct non-trivial extractors for polynomial sources and prove the following, which is a formal version of our main result, i.e., \cref{thm:k-wise-independence-poly-time-extractor-informal}.


\begin{sectheorem}
\label{thm:k-wise-independence-poly-time-extractor}
Let $d, n, k$ be such that $d \le O(\log \log n / \log \log \log n)$ and $k\ge n - \frac{\sqrt{\log n}}{(C\log\log n / d)^{d/2}}$ where $C > 0$ is some large universal constant.
Let $\cX$ be the class of degree $d$ polynomial sources that output $n$ bits and have min-entropy at least $k$. Then we can construct an extractor for $\cX$ in time $\poly(n)$ that extracts $\Omega(\log\log n)$ bits and has error at most $2^{-\Omega(\log \log n)}$. 
\end{sectheorem}



Towards proving the theorem, we first need the following simple observation that the entropy gap of a source cannot get worse by projecting onto a few bits:

\begin{secclaim}
\label{prop:project-source-min-entropy}
Let $\X\sim\F_2^n$ be an arbitrary source such that $\minH(\X) = k$. Let $\X_0$ be projection of $\X$ onto arbitrary $n_0$ bits. Then, $\minH(\X_0) \ge n_0 - (n - k)$.
\end{secclaim}

\begin{proof}
Let $x_0\in \F_2^{n_0}$ be arbitrary and $p_0 = \Pr(\X_0 = x_0)$. Then by an averaging argument, there exists $x\in \F_2^n$ such that the projection of $x$ onto coordinates corresponding to $\X_0$ equals $x_0$. Then, $\Pr(\X = x) \ge p_0\cdot 2^{-(n-n_0)}$. Hence, if $p_0 > 2^{- (n_0 - (n - k))}$, then $\Pr(\X = x) > 2^{-k}$, a contradiction.
\end{proof}

We will use the following lemma to efficiently construct a $t$-wise independent hash function family:
\begin{seclemma}[Follows from \protect{\cite[Corollary 3.3.4]{vadhan12pseudorandomness}}]
\label{lem:k-wise-independence-enumeration}
For every $n, m, t\in \N$, there exists a family of $t$-wise independent functions $\cH = \{h:\zo^n\rightarrow\zo^m\}$ such that we can enumerate the family in time $2^{t\cdot \max(n, m)}\cdot \poly(n, m, t)$ time and evaluate each function in $\poly(n, m, t)$ time.
\end{seclemma}


\RestyleAlgo{ruled}

\begin{algorithm}
\caption{Extractor from $t$-wise independent family}
\SetKwInOut{Input}{input}\SetKwInOut{Output}{output}
\Input{degree $d$, input source length $\ell$, output source length $n_0$, min-entropy $k_0 = n_0 - g$, extractor output length $r$, target error $\varepsilon$, the parameter $t$ for $t$-wise independence}
\Output{An extractor $f$ from $n_0$ bits to $r$ bits with error $\varepsilon$ for degree $d$ polynomial sources from $\ell$ bits to $n_0$ bits if it exists from some $t$-wise independent family}
\label{algo:t-wise-independent-extractor}
Let $\cF$ be some fixed family of $t$-wise independent functions from $n_0$ bits to $r$ bits.\\
\For{every function $f\in \cF$}{
flag $\gets$ True.\\
\For{every degree $d$ polynomial map $\cP$ from $\ell$ bits to $n_0$ bits}{
Brute force over all $2^{\ell}$ assignments to compute min-entropy of $\cP(\U_{\ell})$ and let it be $k_{\cP}$.\\
\If{$k_{\cP} \ge k_0$} {
Brute force over all $2^{\ell}$ assignments to compute $\varepsilon_{f, \cP} = |\U_m - f(\cP(U_{\ell}))|$.\\
\If{$\varepsilon_{f, \cP} > \varepsilon$} {
flag $\gets$ False.
}
}
}
\If{flag $=$ True} {
\Return{f}
}
}
\Return{Fail}
\end{algorithm}

Using these, we prove \cref{algo:t-wise-independent-extractor} yields an extractor for polynomial sources.

\begin{seclemma}\label{lem:t-wise-independent-extractor-constructive-general}
Let $d, g, n, k, r$ be such that $0 \le g \le n, k\ge n-g, d \le \Omega\left(\frac{g}{\log g}\right), O(d + \log g)\le r \le \Omega(g)$.
Let $\cX$ be the class of degree $d$ polynomial sources that output $n$ bits and have min-entropy at least $k$. Then we can construct an extractor for $\cX$ in time $2^{O\left(\binom{\Theta(r)}{\leq d}\cdot g^2\right)}$ that extracts $r$ bits and has error $2^{-\Omega(r)}$. 
\end{seclemma}

\begin{proof}
Let $\X\in \cX$ be arbitrary. Consider the first $n_0 = 1.01g$ bits of $\X$ and let this source be $\X_0$. Then, by \cref{prop:project-source-min-entropy}, it holds that $\minH(\X_0) \ge n_0 - g \ge \Omega(n_0)$.
We use \cref{lem:new-input-reduction-lemma} with min-entropy $k_0 = \Omega(r)$ to infer that it suffices to construct extractors polynomial sources with input length $\ell = \Theta(k_0)$.
By \cref{lem:t-wise-independent-image-extractor-exists}, there exists a function $f:\F_2^{n_0}\rightarrow \F_2^{r}$ such that for all polynomial sources $\Y$, $|f(\Y) - \U_r| \le 2^{-\Theta(k_0)}$. Moreover, such $f$ will be one of the functions in family of $t$-wise independent functions where $t = 2\log(\Theta(k_0) + |\cX|)$. By setting these input parameters to \cref{algo:t-wise-independent-extractor}, we will indeed find such an $f$.

Let us analyze the runtime of \cref{algo:t-wise-independent-extractor}. The number of degree $d$ sources with input length $\ell$ and output length $n_0$ is $2^{\binom{\ell}{\le d}\cdot n_0}$. The time to enumerate the $t$-wise independent family is $2^{tn_0}\poly(t, n_0) \le 2^{2\binom{\ell}{\le d}\cdot n_0^2}\poly(\ell, d, n_0)$ (\cref{lem:k-wise-independence-enumeration}). Computing entropy and checking if the function is an extractor takes $O\left(2^{O(\ell + n_0)}\cdot \poly(n_0)\right)$. As $\ell = \Theta(r)$ and $d \le O(n_0 / \log n_0)$, the overall runtime of this algorithm is $2^{O\left(\binom{\Theta(r)}{\le d}\cdot n_0^2\right)}$. As $n_0 = 1.01g$, the runtime is as desired.
\end{proof}

We specialize lemma \cref{lem:t-wise-independent-extractor-constructive-general} to obtain \cref{thm:k-wise-independence-poly-time-extractor}.

\begin{proof}[Proof of \cref{thm:k-wise-independence-poly-time-extractor}]
Set $g = \frac{\sqrt{\log n}}{(C\log\log n / d)^{d/2}}$ for some large constant $C$, and $r = \Theta(\log g + d\log \log g)$ in \cref{lem:t-wise-independent-extractor-constructive-general}.
\end{proof}

\dobib

%% file: sections/impossibility.tex
\section{Impossibility results}
\label{sec:impossibility}


In this section, we show various impossibility results for polynomial NOBF sources and hence, these results apply to both polynomial sources and variety sources. We first show a sampling result that demonstrates power of the quadratic NOBF sources: they can sample optimal sized Sidon sets. This directly yields an impossibility result against sumset extractors, albeit with weaker parameters than in \cref{thm:quadratic-nobf-sumset-extractor-lower-bound-informal}. We then show affine dispersers cannot be used to disperse from degree $d$ polynomial NOBF sources below certain min-entropy (this is tight). We finally will prove \cref{thm:quadratic-nobf-sumset-extractor-lower-bound-informal}, that sumset extractors cannot be used to even disperse against quadratic NOBF sources below certain min-entropy.

\subsection{A warm-up via Sidon sets}

To start things off, let us recall the definition of Sidon sets.

\begin{secdefinition}[Sidon sets]
We say $S\subseteq \F_2^n$ is a Sidon set if for all $a, b, c, d\in S$ such that $a\ne b, c\ne d, \{a, b\} \ne \{c, d\}$, it holds that $a + b \ne c + d$.
\end{secdefinition}

We show that quadratic NOBF sources can uniformly sample the largest possible Sidon sets over $\F_2^n$, and thus we cannot use sumset extractors below min-entropy $n/2$ to extract from polynomial NOBF sources. Later, we will obtain a much stronger version of the latter claim.

We consider the correspondence between $\F_{2^t}$ and $\F_2^t$:

\begin{secdefinition}
For a finite field $\F_{2^t}$, we define the function $\phi:\F_{2^t}\rightarrow \F_2^t$ that sends the field element to its vector representation.
\end{secdefinition}

We observe that $\phi$ is additive:

\begin{secfact}
\label{fact:vector-rep-additive-bijection}
For all $x, y\in \F_{2^t}$, it holds that $\phi(x+y) = \phi(x) + \phi(y)$. Moreover, $\phi$ is a bijection.
\end{secfact}

We will use the following nice lemma involving $\phi$:

\begin{seclemma}[\protect{\cite[Lemma 2.3.1]{swastikkoppartyphdthesis}}]
\label{lem:multilinear-polynomial-weight-degree}
Let $p:\F_{2^t}\rightarrow \F_{2^t}$ be a degree $d$ polynomial, and let the Hamming weight of $d$ (when expressed in binary) be $w$. Then, there exists a degree $w$ multilinear polynomial $q:\F_2^t\rightarrow \F_2^t$ such that for all $x\in \F_{2^t}$, it holds that $\phi(p(x)) = q(\phi(x))$.
\end{seclemma}

Using these, we show there exists a quadratic NOBF source that uniformly sample largest possible Sidon set over $\F_2^n$:

\begin{secclaim}\label{claim:2-poly-sidon-set}
There exists a degree $2$ polynomial NOBF source $\Y$ with $\minH(\Y) = n/2$ such that $\Y$ uniformly samples a Sidon set.
\end{secclaim}


\begin{proof}
Consider the set $S = \{(x, x^3): x\in \F_{2^{n/2}}\}$. It's well known that this set is a Sidon set \cite{rrw22sidon}.
Let $\phi:\F_{2^{n/2}}\rightarrow \F_2^{n/2}$ be the function that sends the field element to their vector representation.
Using  \cref{lem:multilinear-polynomial-weight-degree}, we infer that there exists a degree $2$ polynomial map $q:\F_2^{n/2}\rightarrow \F_2^{n/2}$ such that for all $x\in \F_2^{n/2}$, $\phi(x^3) = q(\phi(x))$. Applying \cref{fact:vector-rep-additive-bijection}, we infer that $T = \{(y, q(y)): y\in \F_2^{n/2}\}$ is also a Sidon set.
We define $\Y\sim\F_2^n$ to be the degree $2$ polynomial NOBF source that is uniform over the set $T$.
Then, $\Y$ uniformly samples a Sidon set and $\minH(\Y) = n/2$.
\end{proof}

We show the following towards our impossibility result:

\begin{secclaim}
Let $S\subset \F_2^n$ be a Sidon set. Then, for all $A, B\subset \F_2^n, |A| \ge 2, |B| \ge 3: A+B\not\subset S$.
\end{secclaim}

\begin{proof}
Say such $A, B$ existed. Then, pick $a_1, a_2\in A$ and $b_1, b_2\in B$ such that $a_1 + a_2 \ne b_1 + b_2$. Let $C = \{a_1+b_1, a_1 + b_2, a_2+b_1, a_2+b_2\}$. Then, $|C| = 4$ and $C\subset S$. However, $(a_1+b_1) + (a_1+b_2) = (a_2+b_1) + (a_2+b_2)$ which contradicts the fact that $S$ is a Sidon set.
\end{proof}

From these, we infer an impossibility result as a corollary:

\begin{seccorollary}
There exists a degree $2$ polynomial NOBF source $\Y$ with $\minH(\Y) = n/2$ such that for all $A, B\subset \F_2^n, |A| \ge 2, |B| \ge 3: A+B\not\subset \supp(\Y)$.
\end{seccorollary}

Looking ahead, we can apply \cref{lem:worst-case-average-case-sumset} and \cref{lem:convex-comb-far} to infer that we cannot use a sumset extractor in a blackbox way to extract from polynomial NOBF sources of min-entropy $n/2$.

\subsection{Affine dispersers cannot disperse from polynomial NOBF sources}

We show that one cannot use affine dispersers to disperse from degree $d$ polynomial NOBF sources below min-entropy $n - \frac{n}{(\log n)^{d-1}}$.
By \cite[Theorem 5]{cohental15structural}, we know that that an affine disperser for dimension $O(\log n)$ is a disperser for degree $d$ polynomial NOBF sources with min entropy $n - \frac{n d^{d-1}}{(\log n)^{d-1}}$. As polynomial NOBF sources are also variety sources, and affine dispersers below min-entropy $\log n$ cannot exist, this result is indeed tight for any constant $d$.

\begin{sectheorem}
\label{thm:nobf-affine-lower-bound}
Let $c_1 > 0$ be an arbitrary constant. Then, there exists another constant $c_2 > 0$ such that the following holds:
There exists a degree $d$ polynomial NOBF source $\X$ with $\minH(\X) = n - c_2\frac{n}{(\log n)^{d-1}}$ such that $\supp(\X)$ does not contain any affine subspace of dimension $c_1\log n$.
\end{sectheorem}

As affine dispersers with min-entropy requirement $\log n$ can't exist, it indeed follows that we can't use affine dispersers to disperse from polynomial NOBF sources with the stated min-entropy bound.

We first show that there exists a degree $d$ polynomial map that will not become a linear map over any small affine subspace.

\begin{secclaim}
\label{prop:polynomials-restricted-non-linear}
There exists a universal constant $c$ such that the following holds. Let $d, n, t$ be such that $2 \le d < n/2$ and $t < n$. Then, there exist degree $d$ polynomials $p_1, \dots, p_t$ such that on every affine subspace $U$ of dimension $k\ge cd\cdot (n/t)^{1/(d-1)}$, there exists at least one $i$ such that $p_i$ has degree $\ge 2$.
\end{secclaim}

\begin{proof}
We will show such a polynomial map exists using the probabilistic method.
Let $p_1, \dots, p_t:\F_2^n\rightarrow \F_2$ be random polynomials of degree $d$. Let $U$ be arbitrary but fixed affine subspace of dimension $k$. Then, $p_1|_{U}, \dots, p_t|_{U}$ are also uniform polynomials over $k$ variables of degree $d$.
Hence, it must be that:
\[
\Pr_{p_1, \dots, p_t}\left[\bigwedge_{1\le i\le t} \deg(f|_{U}) \le 1\right] \le 2^{-\left(\binom{k}{\le d} - \binom{k}{\le 1}\right)t}
\]
We union bound over all $\le 2^n \binom{2^n}{k}$ affine subspaces of dimension $k$ and see that the probability that there exists some affine subspace of dimension $k$ over which all these polynomials have degree at most $1$ is at most
\[
2^{-\left(\binom{k}{\le d} - \binom{k}{\le 1}\right)t}\cdot 2^n\cdot \binom{2^n}{k}
\]
We set $c$ to a large constant so that the above probability less than $1$.
\end{proof} 

We now show there exists a polynomial NOBF source that does not contain any small affine subspace.

\begin{secclaim}
\label{prop:nobf-affine-lower-bound-general}
There exists a universal constant $c$ such that the following holds: Let $d, k$ be such that $2\le d < k/2$. For any $0 < t < k$, there exists a degree $d$ polynomial NOBF source $\X$ over $k+t$ bits with $\minH(\X) = k$ such that $\supp(\X)$ does not contain any affine subspace of dimension $cd\cdot (k/t)^{1/(d-1)}$.
\end{secclaim}

\begin{proof}
Let $s = cd\cdot (k/t)^{1/(d-1)}$.
Let $(p_1, \dots, p_t): \F_2^k\rightarrow \F_2^t$ be the $t$ polynomials from \cref{prop:polynomials-restricted-non-linear}. Let $\X$ be the polynomial NOBF source over $k+t$ bits where first $k$ bits are $x_1, \dots, x_k$ and last $t$ bits are $p_1(x_1, \dots, x_k), \dots, p_t(x_1, \dots, x_k)$. 

Assume that there exists an affine subspace $U\subset \supp(\X)$ such that $\dim(U) = s$. Observe that once the first $k$ bits of $\X$ are fixed, the last $t$ bits are also fixed. As $U\subset \supp(\X)$, $U$ must also have this property. Let $P\subset \F_2^k$ be the projection of $U$ over the first $k$ bits. Then, $\dim(P) = \dim(U) = s$. Moreover, as $U$ is an affine subspace, the last $t$ bits of $U$ are linear functions of the first $k$ bits. However, this implies that for each $1\le i\le t$, $\deg(p_i|_{P})\le 1$, which is a contradiction.
\end{proof}

\begin{proof}[Proof of \cref{thm:nobf-affine-lower-bound}]
Let $C$ be a large enough constant. We apply \cref{prop:nobf-affine-lower-bound-general} and choose an integer $t$ such that $t \approx k\left(\frac{Cd}{\log (k+t)}\right)^{d-1}$ to infer the claim.
\end{proof}

\subsection{Sumset dispersers cannot disperse from polynomial NOBF sources}
\label{subsec:nobf-lower-bound-sumset-disperser}

We show that we cannot use sumset dispersers to disperse from quadratic NOBF sources below min-entropy $n - n/\log n$.

\begin{sectheorem}
\label{thm:quadratic-nobf-sumset-disperser-lower-bound}
Let $c_1 > 0$ be an arbitrary constant. Then, there exists another constant $c_2 > 0$ such that the following holds:
There exists a degree $2$ polynomial NOBF source $\X$ with $\minH(\X) = n - c_2\frac{n}{\log n}$ such that $\supp(\X)$ does not contain any sumset $A + B$ where $|A| \ge n^{c_1}, |B| \ge n^{c_1}$.
\end{sectheorem}

In fact, we will prove the following, more fine-grained, version of \cref{thm:quadratic-nobf-sumset-disperser-lower-bound}.

\begin{secclaim}
\label{prop:quadratic-nobf-no-sumset-general}
There exists a universal constant $c$ such that the following holds. For any $0 < t < k$, there exists a degree $2$ polynomial NOBF source $\X$ over $n = k+t$ bits with $\minH(\X) = k$ such that $\supp(\X)$ does not contain any sumset $A + B$ where $|A| \ge 2^{cn/t}, |B| \ge 2^{cn/t}$.
\end{secclaim}

Given the above claim, it is easy to prove \cref{thm:quadratic-nobf-sumset-disperser-lower-bound}.

\begin{proof}[Proof of \cref{thm:quadratic-nobf-sumset-disperser-lower-bound}]
The theorem follows by setting $t = O(n/\log n)$ in \cref{prop:quadratic-nobf-no-sumset-general}.
\end{proof}

The rest of this section is devoted to proving \cref{prop:quadratic-nobf-no-sumset-general}. To do so, we prove the following two claims.

\begin{secclaim}
\label{prop:random-poly-map-no-affine-sumset}
There exists a universal constant $c$ such that the following holds. Let $t, n\in \N$ be such that $t < n$. There exist degree $2$ polynomials $p_1, \dots, p_t:\F_2^n\rightarrow \F_2$ such that for every pair of affine subspaces $U, V$ of dimensions $r \ge cn / t$ each, and for all $y\in \F_2^n$, there exists at least one $i$ and at least one $u\in U, v\in V$ such that $p_i(u+v) \ne p_i(u) + p_i(v) + y_i$.
\end{secclaim}

\begin{secclaim}
\label{prop:poly-map-sumset-is-affine}
Let $P = (p_1, \dots, p_t): \F_2^n\rightarrow \F_2^t$ be a degree $2$ polynomial map. Let $y\in \F_2^t$ be arbitrary. Let $A, B\subset \F_2^n$ be such that for all $a\in A, b\in B$, it holds that $P(a) + P(b) = P(a+b) + y$. Then, there exist affine subspaces $U, V\subset \F_2^n$ such that for all $u\in U, v\in V$, it holds that $P(u) + P(v) = P(u+v) + y$ and $|U| \ge |A|, |V| \ge |B|$.
\end{secclaim}

Using these claims, we can indeed get a general trade-off for quadratic NOBF sources not containing any sumset:

\begin{proof}[Proof of \cref{prop:quadratic-nobf-no-sumset-general}]
Let $\X$ be the polynomial NOBF source where the first $k$ bits are independent and uniform and the last $t$ bits are outputs of polynomial map $P$ from \cref{prop:random-poly-map-no-affine-sumset} with input variables as the first $k$ bits (set $c$ to the universal constant from there).
We now proceed by contradiction and assume there exist $A, B\subset \F_2^n$ such that $|A| \ge 2^{cn/t}, |B| \ge 2^{cn/t}$, and $A+B\subset \supp(\X)$.
Let $a_0\in A, b_0\in B$ be arbitrary. Let $A^\pr = a_0+A, B^\pr = b_0+B, \X^\pr = \X + (a_0 + b_0)$. Then, $A^\pr+B^\pr\subset \supp(\X^\pr)$.
Observe that $0^n\in A^\pr$ and $0^n\in B^\pr$. So, $A^\pr\subset \supp(\X^\pr)$, and  $B^\pr\subset \supp(\X^\pr)$.
Moreover, $\X^\pr$ is a degree $2$ polynomial NOBF source with $\minH(\X^\pr) = \minH(\X)$.

Let the last $n-k$ bits of $\X^\pr$ be the output of the degree $2$ polynomial map $P^\pr$.
Let $A^{\pr}_0, B^{\pr}_0 \subset \F_2^k$ be the projections of $A^{\pr}, B^{\pr}$ respectively onto the first $k$ bits.
As $A^{\pr}\subset \supp(\X^{\pr}), B^{\pr}\subset \supp(\X^{\pr})$, and the last $n-k$ bits are deterministic functions of the first $k$ bits, it must be that $|A^{\pr}_0| = |A^{\pr}|$ and $|B^{\pr}_0| = |B^{\pr}|$.
Similarly, as $A^\pr+B^\pr\subset \supp(\X^\pr)$, it must be that $P^\pr(A^{\pr}_0) + P^\pr(B^{\pr}_0) = P^\pr(A^{\pr}_0 + B^{\pr}_0)$.
By \cref{prop:poly-map-sumset-is-affine}, there exist affine subspaces $U^\pr, V^\pr\subset \F_2^k$ such that for all $u^{\pr}\in U^{\pr}, v^{\pr}\in V^{\pr}$, it holds that $P^{\pr}(u^{\pr}) + P^{\pr}(v^{\pr}) = P^{\pr}(u^{\pr} + v^{\pr})$ where $|U^{\pr}| \ge |A^{\pr}_0| = |A^{\pr}|, |V^{\pr}| \ge |B^{\pr}_0| = |B^{\pr}|$.

Observe that $P^\pr(x) = P(g+x) + h$ where $g\in \F_2^k, h\in \F_2^t$ are some fixed strings.
Then, $P(g+U^{\pr}) + P(g+V^{\pr}) = P(g+U^{\pr}+V^{\pr}) + h$.
Let $U, V\subset \F_2^k$ be such that $U = g+U^{\pr}, V = g+V^{\pr}$.
Then $U, V$ are affine subspaces, $P(U) + P(V) = P(U+V) + h$, and $|U| = |U^{\pr}| \ge |A^{\pr}| = |A|, |V| = |V^{\pr}| \ge |B^{\pr}| = |B|$. 
However, this is a contradiction to the choice of $P$.
\end{proof}

We now prove the helpful claim that there exists a quadratic map $P$ that has the property that for every pair of high dimension affine subspaces $(U, V)$, there exist $u\in U, v\in V$ such that $P(u) + P(v) \ne P(u+v)$.

\begin{proof}[Proof of \cref{prop:random-poly-map-no-affine-sumset}]
We will show such a map exists using the probabilistic method.
Let $P = (p_1, \dots, p_t):\F_2^n\rightarrow \F_2^t$ be a random degree $2$ polynomial map.
Fix $y\in \F_2^t$. At the end, we will union bound over these $2^t$ distinct $y$.
Without loss of generality assume that $y = 0^t$.
Indeed, we can define $P^{\pr}(x) = P(x) + y$
so that for all $a\in A, b\in B$, it holds that $P^{\pr}(a) + P^{\pr}(b) = P^{\pr}(a+b)$.
Moreover, if there exist $U, V$ such that for all $u\in U, v\in V$: $P^{\pr}(u) + P^{\pr}(v) = P^{\pr}(u+v)$, then we will recover that for all $u\in U, v\in V$: $P(u) + P(v) = P(u+v) + y$ as desired.
Moreover, $P^{\pr}$ as defined will be distributed as a uniformly random degree $2$ polynomial map.

Call a degree $2$ polynomial map $P$ `bad' if there exist affine subspaces $C, D$ of dimensions $r$ each such that for all $c\in C, d\in D$ it holds that $P(c) + P(d) = P(c + d)$.
Call such $C, D$ the affine subspaces that `witness' the badness of $P$.
We will go over each pair of affine subspaces $C, D$ and show that the fraction of bad maps witnessed by the pair $(C, D)$ are very small.

Let $U, V$ be arbitrary dimension $r$ subspaces.
Let $u_0, v_0\in \F_2^n$ be arbitrary.
Then, we fix pair of dimension $r$ affine subspaces $(u_0 + U, v_0 + V)$ and bound the fraction of bad maps witnessed by the pair.
We consider two cases:
\begin{casesenum}
    \item $\dim(U\cap V) \ge r/2$.\\
    Let $W = (U\cap V)$.
    Let $P$ be a bad map witnessed by $u_0 + U, v_0 + V$, i.e., for all $u\in (u_0 + U), v\in (v_0 + V)$, it holds that $P(u) + P(v) = P(u + v)$.
    We claim that $P|_{(u_0 + W)}$ is a degree $1$ polynomial map.
    Indeed, above condition guarantees that $\forall w_1, w_2\in W: P(u_0 + w_1) + P(v_0 + w_2) = P(u_0 + v_0 + w_1 + w_2)$.
    This also implies that $\forall w\in W: P(u_0 + w) + P(v_0 + w) = P(u_0 + v_0)$.
    Repeatedly applying these, we infer that:
    \begin{align*}
    P(u_0 + (w_1 + w_2))
    & = P(u_0 + v_0) + P(v_0 + (w_1 + w_2))\\
    & = P(u_0 + v_0) + (P(u_0 + w_1) + P(u_0 + v_0 + w_2))\\
    & = P(u_0 + w_1) + P(u_0 + v_0) + P(v_0 + (u_0 + w_2))\\
    & = P(u_0 + w_1) + (P(u_0) + P(v_0)) + (P(v_0) + P(u_0 + w_2)))\\
    & = P(u_0 + w_1) + P(u_0 + w_2) + P(u_0)\\
    \end{align*}
    Hence, $P$ restricted to $u_0 + W$ is indeed an affine map.
    We observe that $p_1|_{u_0 + W}, \dots, p_t|_{u_0+W}$ are distributed as uniform degree at most $2$ polynomials over $r/2$ variables.
    The probability that each of these polynomials has degree at most $1$ is at most $2^{-\binom{r/2}{2}t}$.

    \item $\dim(U\cap V) < r/2$.\\
    Let $u_0 + S$ be the largest affine subspace inside $u_0 + U$ such that $S\cap (U\cap V) = \{0\}$.
    Similarly, let $v_0 + T$ be the largest affine subspace inside $V$ such that $T\cap (U\cap V) = \{0\}$. It must be that $\dim(S), \dim(T) \ge r/2$ and $S\cap T = \{0\}$. By considering appropriate subsets of $S$ and $T$, we without loss of generality assume $\dim(S) = \dim(T) = r/3, (u_0 + S)\cap (T\cup (v_0 + T)) = (v_0 + T)\cap (S\cup (u_0 + S)) = \emptyset$. 
    Let basis vectors of $S$ and $T$ be $(s_1, \dots, s_{r/3})$, and $(t_1, \dots, t_{r/3})$ respectively. Without loss of generality, let it be that $u_0+s_1, \dots, u_0+s_{r/3}$ are linearly independent and $v_0 + t_1, \dots, v_0 + t_{r/3}$ are also linearly independent.
    Then, by using the various empty intersection conditions above, the vectors $u_0 + s_1, \dots, u_0 + s_{r/3}, v_0 + t_1, \dots, v_0 + r_{r/3}$ are also linearly independent.
    Let $b_1, \dots, b_{n-r/3}$ be linearly independent vectors such that $u_0 + s_1, \dots, u_0 + s_{r/3}, v_0 + t_1, \dots, v_0 + r_{r/3}, b_1, \dots, b_{n-r/3}$ are all linearly independent. Let's rename these vectors to be $c_1, \dots, c_n$.
    
    Now, we choose the random quadratic polynomials $p_1, \dots, p_t$ by randomly sampling monomials of degree at most $2$ over these $c_i$. As the $c_i$ are linearly independent, $P$ will still be a uniformly random quadratic map.
    Let $P$ be a bad map witnessed by $u_0 + U, v_0 + V$.
    We claim there does not exist $i$ such that $p_i$ contains the monomial $c_jc_k$ where $c_j\in \{u_0+s_1, \dots, u_0 + s_{r/3}\}$ and $c_k\in \{v_0 + t_1, \dots, v_0 + t_{r/3}\}$.
    By way of contradiction, assume there exists $i$ such that $p_i$ contains the monomial $c_jc_k$ where $c_j\in \{u_0+s_1, \dots, u_0 + s_{r/3}\}$ and $c_k\in \{v_0 + t_1, \dots, v_0 + t_{r/3}\}$.
    Without loss of generality we assume that the singleton monomials $c_j$ and $c_k$ are not present in $p_i$ and the degree $0$ monomial is also absent (if any of them are present, then we can easily change assignments $\alpha_1, \alpha_2$ and their outcomes below and get the same claim).
    Consider the following two assignments:
    \begin{enumerate}
        \item Assignment $\alpha_1$ where $c_j = 1$, and remaining variables are set to $0$.
        \item Assignment $\alpha_2$ where $c_k = 1$, and remaining variables set to $0$. 
    \end{enumerate}
    Then, $p_i(\alpha_1) = p_i(\alpha_2) = 0$.
    As monomial $c_jc_k$ is present in $p_i$, it must be that $p_i(\alpha_1 + \alpha_2) = 1$.
    However, this implies we found $\alpha_1\in (u_0 + U)$ and $\alpha_2\in (v_0 + V)$ such that 
    $P(\alpha_1) + P(\alpha_2) \ne P(\alpha_1 + \alpha_2)$, contradicting the fact that $u_0 + U, v_0 + V$ witnessed badness of $P$.
    Hence, for this not to happen, all such `cross' monomials must not occur in any $p_i$.
    As each $p_i$ is a random quadratic polynomial, the probability that all `cross' monomials are absent in all $p_i$ is at most $2^{-(r/3)^2 t}$.
\end{casesenum}

We union bound over all pairs of affine subspaces of dimension $r$ and consider whether they fall into the first case or the second case. If they fall into the first case, then we only union bound over $\le 2^n\cdot \binom{2^n}{r/2}$ affine subspaces of dimension $r/2$ and consider the probability that a bad map $P$ that they witness becomes linear over that affine subspace. If they fall into the second case, then we union bound over all $\le \left(2^n\cdot \binom{2^n}{r/3}\right)^2$ disjoint pairs of affine subspaces of dimension $r/3$ and consider the probability that any bad map they witness won't have such cross monomials. We finally add both these probabilities to get our final bound. For the first case, the expression will be
\[
2^{-\binom{r/2}{2} t}\cdot 2^n\cdot \binom{2^n}{r/2}
\]
For the second case, the expression will be:
\[
2^{-(r/3)^2 t}\cdot \left(2^n\cdot \binom{2^n}{r/3}\right)^2
\]
We set $r = cn/t$ where $c$ is a large universal constant so that the sum of the above probabilities is much smaller than than $2^{-t}$. Finally, we union bound over all $2^t$ of the $y\in \F_2^t$ to get the desired claim.
\end{proof}

We lastly prove the useful claim that for a quadratic map $P$, if there exist sets $A, B$ such that for all $a\in A, b\in B$ it holds that $P(a) + P(b) = P(a+b)$, then we can also find affine subspaces $U, V$ with the same property and of larger sizes. We first need the notion of directional derivatives:

\begin{secdefinition}[Directional derivative]
For a polynomial $p:\F_2^n\rightarrow \F_2$ and $a\in \F_2^n$, we define its directional derivative in direction $a$, i.e., $D_a(p)(\cdot):\F_2^n\rightarrow\F_2$ as
\[
    D_a(p)(x) = p(x) + p(x+a)
\]
\end{secdefinition}
Clearly $D_a(p)$ is a polynomial. It's well known that $\deg(D_a(p)) \le \deg(p)-1$. 
We also extend the definition of directional derivatives to apply to a polynomial map $P:\F_2^n\rightarrow\F_2^t$.
For a fixed direction $a\in \F_2^n$, we define $D_a(P)(\cdot):\F_2^n\rightarrow\F_2^m$ as $D_a(P)(x) = (D_a(p_1)(x), \dots, D_a(p_m)(x))$. Using these, we present our proof:

\begin{proof}[Proof of \cref{prop:poly-map-sumset-is-affine}]
Without loss of generality assume that $y = 0^t$. Indeed, let $P^{\pr}(x) = P(x) + y$ so that for all $a\in A, b\in B$: $P^{\pr}(a) + P^{\pr}(b) = P^{\pr}(a+b)$.
Moreover, if there exist such affine subspaces $U, V$ so that for all $u\in U, v\in V$: $P^{\pr}(u) + P^{\pr}(v) = P^{\pr}(u+v)$, then we will indeed recover the fact that for all $u\in U, v\in V$: $P(u) + P(v) = P(u+v) + y$ as desired.
Let $C, D$ be such that $A\subset C$, $B\subset D,$ and for all $c\in C, d\in D$ it holds that $P(c) + P(d) = P(c+d)$. Moreover, let $C$ and $D$ be the largest such sets. 
To prove the claim, it suffices to show that $C$ and $D$ are affine subspaces.

For $a\in \F_2^n$, define $D_a(P)(x) = (D_a(p_1)(x), \dots, D_a(p_t)(x)) = P(x) + P(x+a)$, the map of directional derivatives in direction $a$. Let $S_a = \{z\in \F_2^n: D_a(P)(z) = P(a)\}$. We claim that $y\in S_a \iff P(y) + P(a) = P(a+y)$. Indeed,
\[
  D_a(P)(y) = P(a) \iff  P(y) + P(y+a) = P(a) 
\]
Let $S_C = \bigcap_{c\in C} S_c$. Then, it must be that $B\subset S_C$.
As $D$ is maximal such set and for all $c\in C, s\in S_C$: $P(c) + P(s) = P(c + s)$, it must be that $D = S_C$.
By a symmetric argument, $C = S_D$. Observe that for arbitrary $a\in \F_2^n$, $S_a$ is an affine subspace. As intersection of affine subspaces is an affine subspace, $S_C = D$ as well as $S_D = C$ are affine subspaces as desired.
\end{proof}

\subsection{Sumset extractors cannot disperse from polynomial NOBF sources}
\label{subsec:nobf-lower-bound-sumset-extractor}

We show that we cannot use sumset extractors to disperse from quadratic NOBF sources below min-entropy $n - n / \log\log n$.

\begin{sectheorem}
\label{thm:quadratic-nobf-sumset-extractor-lower-bound}
Let $0 < \varepsilon < 1, 0 < c_1$ be arbitrary constants. Then, there exists another constant $c_2 > 0$ such that the following holds:
There exists a degree $2$ polynomial NOBF source $\X$ with $\minH(\X) = n - c_2\frac{n}{\log \log n}$ such that $\X$ is $(1-\varepsilon)$-far from a convex combination of sumset sources of min-entropy $c_1 \log n$.
\end{sectheorem}

Note that if a distribution is distance $\frac{1}{2}$ away from any convex combination of sumset sources, then a sumset extractor cannot be used in a blackbox way as a disperser for that distribution. Also, as no sumset extractor can exist for min-entropy below $\log n$, this result indeed shows that we can't use sumset extractors in a blackbox way to disperse from degree $2$ polynomial NOBF sources. 

We first prove a worst case to average case type reduction for sumsets.

\begin{seclemma}
\label{lem:worst-case-average-case-sumset}
Let $0 < \delta < 1$ be a fixed constant. Let $\X\sim\F_2^n$ be such that for all flat sources $\A, \B\sim \F_2^n$ with $\minH(\A) = \minH(\B) = t$, it holds that $(\A+\B)\not\subset \supp(\X)$. Then, for all flat sources $\RR, \SSS\sim\F_2^n$ such that $\minH(\RR) = \minH(\SSS) = c\cdot 2^t$, it holds that $\Pr_{r\sim\RR, s\sim\SSS}[r+s\in \supp(\X)] \le \delta$. Here, $c > 0$ is a constant depending only on $\delta$.
\end{seclemma}

We will show that if a source is far from all sumset sources, then it is also far from all convex combination of sumset sources:

\begin{seclemma}[\protect{Similar to \cite[Theorem 14]{alrabiah2022low}}]
\label{lem:convex-comb-far}
Let $0 \le \delta \le 1, 0\le k$, and $\X\sim \F_2^n$ be such that for all flat sources $\RR, \SSS\sim\F_2^n$ with $\minH(\RR), \minH(\SSS) \ge k$, it holds that $\Pr_{r\sim\RR, s\sim\SSS}[r+s\in \supp(\X)] \le \delta$.
Then, for all $\Y$ such that $\Y$ is a convex combination of sumset sources of min-entropy at least $k$, it holds that $|\X - \Y| \ge 1 - \delta$.
\end{seclemma}

Using these, and \cref{prop:quadratic-nobf-no-sumset-general} from \cref{subsec:nobf-lower-bound-sumset-disperser}, we show that sumset extractors cannot even \emph{disperse} from degree $2$ polynomial NOBF sources:

\begin{proof}[Proof of \cref{thm:quadratic-nobf-sumset-extractor-lower-bound}]
Let $\X$ be source guaranteed by \cref{prop:quadratic-nobf-no-sumset-general} with $\minH(\X) \ge n - c_2\frac{n}{\log \log n}$ such that for all $A, B\subset \F_2^n$ with $|A| = |B| = c_3\log n$, it holds that $(A+B)\not\subset \supp(\X)$ (we specify $c_3$ later, depending on $c_1$).
Using \cref{lem:worst-case-average-case-sumset} and by choice of $c_3$, we infer that for all flat sources $\RR, \SSS\sim\F_2^n$ with $\minH(\RR) = \minH(\SSS) = c_1\log n$, it holds that $\Pr(\RR+\SSS)\in \supp(\X) \le \eps$.
Let $\Y\sim\F_2^n$ be arbitrary convex combination of sumset sources $\{(\RR^{(i)} + \SSS^{(i)})\}_i$, each with min-entropy $c_1\log n$.
Applying \cref{lem:convex-comb-far}, we infer that $|\X - \Y| \ge 1 - \eps$ as desired.
\end{proof}

We use a bipartite Ramsey bound to get a worst case to average case type reduction for sumset sources:

\begin{seclemma}[\cite{znam1963bipartite}]
The maximum number of edges in a bipartite graph over $[n]\times [n]$ without inducing a complete bipartite $t\times t$ subgraph is at most $(t-1)^{1/t}\cdot n^{2-1/t} + \frac{1}{2}\cdot (t-1)\cdot n$.
\end{seclemma}

We will utilize the following corollary of this statement:

\begin{seccorollary}
\label{lem:bipartite-ramsey-bound}
Fix $0 < \delta \le 1$.
Let $G$ be a bipartite graph over $[n]\times [n]$ with at $\delta \cdot n^2$ edges.
Then, $G$ induces a complete bipartite subgraph $H$ over $[\varepsilon\cdot \log n]\times [\varepsilon\cdot \log n]$ where $0 < \varepsilon$ is a constant depending only on $\delta$.
\end{seccorollary}

Equipped with this, we prove our main lemma:

\begin{proof}[Proof of \cref{lem:worst-case-average-case-sumset}]
Assume this is not the case and there exist such $\RR$ and $\SSS$. Let $\minH(\RR) = \minH(\SSS) = k$.  Consider a bipartite graph $G$ over $\supp(\RR)\times \supp(\SSS)$ with an edge between $r\in \supp(\RR)$ and $s\in \supp(\SSS)$ if $r+s \in \supp(\X)$.
By assumption, $G$ has at least $\delta\cdot 2^{2k}$ edges.
Using \cref{lem:bipartite-ramsey-bound}, we infer that $G$ induces a complete bipartite subgraph where where each part has size $\varepsilon \cdot k$ ($\varepsilon$ depends only on $\delta$). Equivalently, there exist sets $C\subset \supp(\RR), D \subset \supp(\SSS)$ such that $|C| = |D| = \varepsilon\cdot k$ and $(C+D) \subset \supp(\X)$. Let $\A$ be the uniform distribution over $C$ and $\B$ be the uniform distribution over $D$. Then, $\minH(\A) = \minH(\B) = \log(\varepsilon\cdot k)$. Setting $c = 1/\varepsilon$, we get a contradiction.
\end{proof}

Finally, we show that if a distribution is far from every sumset source, then it's far from every convex combination of sumset sources.

\begin{proof}[Proof of \cref{lem:convex-comb-far}]
Let $\Y\sim\F_2^n$ be arbitrary convex combination of sumset sources $\{(\RR^{(i)} + \SSS^{(i)})\}_i$, each with min-entropy at least $k$. Let $T = \supp(\X)$.
Then,
\begin{align*}
|\X-\Y| & \ge \Pr[\Y\in \overline{T}] -  \Pr[\X\in \overline{T}] = \Pr[\Y\in \overline{T}]\\
& \ge \min_i \Pr[\Y^{(i)}\in \overline{T}] = 1 - \max_i \Pr[\Y^{(i)}\in T]\\
& \ge 1 - \delta.
\end{align*}
\end{proof}

\dobib

%% file: sections/open.tex
\section{Open problems}
\label{sec:open}

The problem of constructing extractors for sources sampled by $\mathbb{F}_2$-polynomials is a natural one, and we view our results as  initial progress on this question. We leave open a   number of interesting open directions:

\begin{enumerate}
    \item 
     Construct extractors or dispersers for polynomial sources with better min-entropy dependence than what we constructed here. For instance, some interesting potential candidates to explore are the MAJORITY function, or the generalized inner product function.

    \item 
    It will be interesting to make progress on the easier question of extracting from constant degree polynomial NOBF sources below min-entropy $0.999n$. Extracting from constant degree variety sources below min-entropy $0.999n$ is an important open problem and here, we introduced an interesting subclass of variety sources - polynomial NOBF sources - for which we also don't have better extractors.
    
    An even simpler question is to construct \emph{dispersers} for constant degree polynomial NOBF sources below min-entropy $n/2$. Note that for any NOBF source with $> n/2$ good bits, the MAJORITY function is a disperser.

    \item 
    Construct extractors or dispersers for polynomial sources with degree $\poly(\log n)$. Such an extractor will also extract from sources sampled by $\AC^0[\oplus]$ circuits, a model for which no non-trivial extractors are known. Our constructions work for degree up to $O(\log \log n)$, and thus fall short of achieving this.
\end{enumerate}



\dobib

%% file: sections/acknowledgements.tex
\section{Acknowledgements}

We want to thank Michael Jaber for helpful discussions.

\dobib